\documentstyle[12pt,epsfig]{article}
\begin{document}
\thispagestyle{empty}
 \centerline{EUROPEAN ORGANIZATION FOR NUCLEAR RESEARCH}
  \vspace{5mm}

    \rightline{SL / Note 93-74 (AP)}
       \leftline{VB/jt}
    \vspace{35mm}
 \centerline{\LARGE\bf Crystal Channelling Simulation}
\vspace{.5cm}
\centerline{\LARGE\bf  CATCH 1.4 User's Guide}
\vspace{20mm}

\centerline{\Large\bf Valery Biryukov\footnote{Permanent address:
IHEP, Protvino, 142284 Moscow Region, Russia}}
 \vspace{5mm}

 \centerline{\large\bf CERN, Geneva, Switzerland}
 \vspace{20mm}

\centerline{\small\bf Abstract}
 \vspace{5mm}

{The Monte Carlo program CATCH (Capture And Transport of CHarged
particles in a crystal) for the simulation of planar channelling
in bent crystals is presented. The program tracks a charged
particle through the distorted-crystal lattice with  the use of
continuous-potential approximation and the non-diffusion approach
to the processes of scattering on electrons and nuclei.  The
output consists of the exit angular distributions, the energy
loss spectra, and the  spectra of any close-encounter process of
interest.  The curvature variability, face twist, and various
surface imperfections of the real crystal can be taken into
account.  }
\vspace{10mm}

\centerline{\normalsize Geneva, Switzerland}
\vspace{2mm}

\centerline{\normalsize 13 July 1993}

\pagebreak\thispagestyle{empty}\tableofcontents\listoffigures
\pagebreak\setcounter{page}{1}

\section{ Introduction}
Channelling of a beam of charged particles in a bent monocrystal, is
going to become a working tool for the next generation of accelerators
\cite{b}.  Therefore there is a need for a theoretical tool which
describes a whole set of experimental data on channelling in the GeV
range,  and  which also simulates the processes important for the future
applications.

The output should be the distribution of exiting primary and secondary
particles, the energy loss in  crystal, and any other interesting
quantity related to channelling.  Since these processes are sensitive to
the orientation, the simulation should track every particle through the
crystal lattice computing the probability of any process as a function
 of the co-ordinates.

Historically, one of the independent discoveries of channelling, in the
early sixties, was done in a Monte Carlo simulation of low-energy
($\le$MeV) ions propagating in crystals \cite{r}.  The very low ranges
and the thin crystals used allowed the study of binary collisions of the
incident ion with the atoms of the crystal.  At GeV energy , crystals of
a few centimetres in length are used, therefore tracking with binary
collisions would take a considerable time. Instead, an approach with
continuous potential introduced by J.~Lindhard \cite{l} can be used.  In
 this approach one considers collisions of the incoming particle with
the atomic strings or planes instead of separate atoms, if the particle
is sufficiently aligned with respect to the crystallographic axis or
plane.  The validity of doing this improves with the increase of the
particle velocity \cite{l}.

Besides the motion in the potential one must take into account the
scattering.  This makes it necessary to use either  kinetic equations
\cite{ki}, or computer simulation \cite{mc} to transport particles
through a crystal.  The general feature of the methods described in
Refs. \cite{ki,mc} is the use of the diffusion approach which omits the
single scattering acts.  However, in Monte Carlo methods it is easy to
include the single collisions with nuclei and electrons. Besides their
influence on channelling, these close-encounter processes are the
 source of secondary particles emitted from the crystal.  Moreover, such
 collisions with electrons provide interesting peculiarities of the
energy loss spectra in aligned crystals.  Here we describe the Monte
Carlo program  CATCH for the simulation of  planar channelling, which
does not use the diffusion approximation.

To simulate the nuclear collisions we use the LUND \cite{lund} routines;
 the CATCH program serves as a frame to provide the orientational
  dependence of these processes.  Another useful feature of the CATCH
program is  the various imperfections of the crystal surface,
incorporated in the simulation because of their essential role in the
crystal-assisted beam extraction.  Version 1.4 of CATCH allows variable
curvature of the crystal, both longitudinal and transversal, in order to
 set, for instance,  a twist of crystal faces.

\section{Continuous potential} For the potential of the atomic plane we
use the Moli\`{e}re approximation; details can be found e.g. in the
review by D.~Gemmel \cite{g}. The static-lattice potential is modified
 to take into account the thermal vibrations of the lattice atoms; this
is done by integration over the Gaussian distribution of the atom
 displacement.  Bending of the crystal has no effect on this potential.
However, it causes a centrifugal force in the non-inertial frame related
to the atomic planes.  To solve the equation of motion in the potential
$U(x)$ of the bent crystal, as a first approximation to the transport of
a particle,

\begin{equation}
pv \frac{d^2x}{dz^2} = - \frac{dU(x)}{dx} - \frac{pv}{R(z)}\,,
\end{equation}
($x$ being the transversal, $z$ the longitudinal coordinate, $pv$
the particle longitudinal momentum and velocity product, $R(z) $
the local radius of curvature), we use the fast form of the Verlet
algorithm \cite{tob}:
\begin{equation}
x_{i+1} - x_i = (\theta_i + 0.5 f_i\delta z)\delta z  \, ,
\end{equation}
\begin{equation}
\theta_{i+1} - \theta_i =  0.5 (f_{i+1} + f_i)\delta z
\end{equation}
with $\theta $ for $dx/dz$, $f$ for the `force', and $\delta z$ for the
step.  It was chosen over the other second order algorithms for
non-linear equations of motion, such as Euler-Cromer's and Beeman's,
owing  to the better conservation of the transverse energy shown in the
potential motion.

\begin{figure}
\begin{center}
\epsfig{file=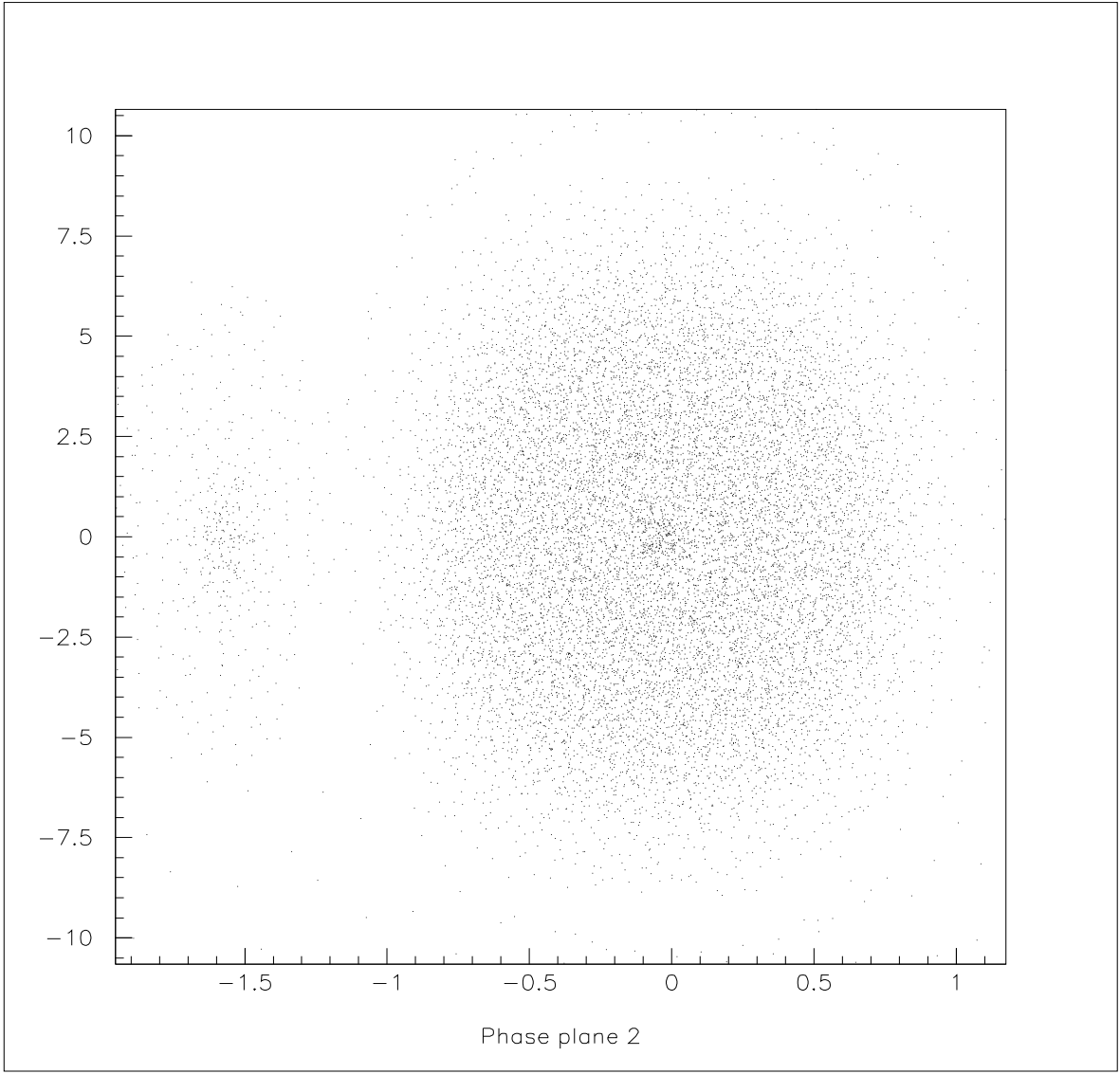,width=\linewidth,height=15cm}     
 \caption{
Phase plane angle ($\mu$rad) vs phase plane position (\AA)
of 450 GeV/c protons traced in Si(111).
 }\label{ph}
\end{center}
\end{figure}

Figure \ref{ph} shows the simulated phase trajectories of protons on the
plane $(x, \theta)$ in Si (111). Scattering was also included; as a
result, there are no channelled protons near the atomic planes.  With
version 1.4 of CATCH  one can specify in the input cards the planar
geometry equivalent  either to  the Si(110) geometry (equidistant
planes), or to the Si(111) one.  It is easy to `build' any other
geometry by editing the program.  The bending curvature  and the crystal
plane orientation can be arbitrary functions of spacial coordinates.
 Any data measured (or assumed) for the real crystal shape can be
implemented in the simulation.

\section{ Scattering} Beam bending by a crystal is due to the trapping
of some particles in the potential well $U(x)$, where they then  follow
the direction of the atomic planes.  This simple picture is disturbed by
scattering processes which could cause (as result of one or many acts)
the trapped particle to come  to a free state (feed out, or de
channelling process), and  an initially free particle to be trapped in
the channelled state (feed in, or volume capture).

\subsection{ Scattering on electrons}
Feed out is mostly due to scattering on electrons \cite{ki},
because the channelled particles keep far from the nuclei.  The
mean energy loss in this scattering can be written as follows
\cite{esb}:
\begin{equation}
- \frac{dE}{dz} = \frac{D}{\beta^2}
(0.5 \ln \frac{2m_ec^2 \beta^2 \gamma^2}{I} -
\beta^2 -\frac{\delta}{2} +
0.5 \rho_e(x) \ln \frac{T_{max}}{I} ) \, ,
\end{equation}
with $D=4\pi N_A r_e^2 m_e c^2 z^2 \frac{Z}{A}\rho$, $z$ for the
charge of the incident particle (in units of $e$), $\rho$ for the
crystal density, $Z$ and $A$ for atomic number and weight, and the
other notation being standard \cite{esb}.

The second part in the brackets is due to  single
 collisions and depends on the local density $\rho_{e}(x)$
(normalized on the amorphous one) of electrons. The angle of
scattering in soft acts can be computed as a random Gaussian with
 r.m.s. value
\begin{equation}
 \theta^2_{rms} = \frac{m_e}{p^2} (\delta E)_{soft} \, ,
\end{equation}
where  $(\delta E)_{soft}$ is the  soft acts, contribution in  Eq.
(4).  The probability of the hard collision contributing to the
second part of Eq. (4) is computed at every step. The energy
transfer  $T$ in such an act is generated according to the
distribution function $P(T)$:
\begin{equation}
 P(T) = \frac{D \rho_e(x)}{2 \beta^2} \frac{1}{T^{2}} \, .
\end{equation}
The transverse momentum transfer $q$ is equal to
\begin{equation}
q = \sqrt{2m_e T + (T/c)^{2}} \, .
\end{equation}
Its projections are used to modify the angles $\theta_x$ and
$\theta_y$ of the particle.

\subsection{ Scattering on nuclei}
The multiple Coulomb scattering on nuclei is computed by the
approximation Kitagawa--Ohtsuki \cite{kit}:
\begin{equation}
\langle \theta^2_{nucl,sc}\rangle \; = \;
\langle \theta^2_{sc}\rangle_{amorph} \cdot \rho_n(x) \,
\end{equation}
i.e. the mean angle of scattering squared is proportional to the
local density of nuclei $\rho_n(x)$.  The density function is
Gaussian with r.m.s. value $u$ being the amplitude  thermal
vibration of the atom.

The probability of nuclear collision, proportional to $\rho_n(x)$,
is checked at every step. If  such a collision succeeds (then the
flag ISTATU is set to --1), the LUND routine responsible for the
event generation may be called.  This scheme may be used for the
description of any close-encounter process of interest; then one
should define the collision length (ADLI in the input cards)
properly.  In version 1.4 of CATCH the secondary particles are
normally not tracked through the crystal lattice, but their exit
angles can be modified due to multiple scattering in the rest part
of the crystal.  In principle, a full-scale tracking of secondary
particles (and even {\em their} products) is possible via the
construction of the appropriate main program (description below)
with loops and trees.

\section{ Crystal imperfection } The complete description of the
program usage follows below.  Here we show how the crystal
imperfection is implemented.
\subsection{ Variable curvature and twist }

If KRADV = 0 is set in the input cards, the transverse curvature
is set to 0.  The longitudinal curvature is 0 for the longitudinal
coordinate $z$ between 0 and STRAIT (length of the forthcoming
straight part), and is constant equal to 1/RADIUS for $z$ between
STRAIT and STRAIT + DLINA (hence DLINA means the length of the
bent part ).

If KRADV = 1 is set in the input cards, the curvature at any point
must be defined by the user in function CURVAT. The total length
of the crystal is STRAIT + DLINA in any case.

If KRADVY = 0 is set in the input cards, the orientation of the
crystal planes  at both entry and exit faces is constant.  If
KRADVY = 1 is set in the input cards, these orientations are
computed according to the functions TCFROY and TCENDY, for the
front face and for the back face respectively.  These functions
must be supplied by the  user.

\subsection{ Surface effects}
In the case of beam extraction from the accelerator, extremely
small impact parameters are possible, making the surface effects
essential.  The particle entering the crystal very close to its
edge, can suffer from various additional factors:
\begin{itemize}
\item miscut angle (between the atomic planes and the surface),
\item roughness (i.e. non flatness) of the surface, \item possible
amorphous layer, \item bent surface.
\end{itemize}

Therefore one must pay particular attention to the near-surface
tracking, where the particule is entering and leaving the cristal
materi al (due to the roughness, holes and bend), both coherent
and non-coherent scattering in this peculiar region, bending in
 short channels, and so on.  The surface effects mentioned above
are simulated in the current version  of CATCH.  The roughness is
expressed by a periodical function, $a \sin(z/\lambda)$, where $a$
is the amplitude of `bumps' and $\lambda$ is their periodicity.
The `rough' crystalline material can be superimposed by a uniform
amorphous layer.  The position of surfaces is computed at every
step in accordance with bending with the variable curvature.  To
set up these imperfections, simply give non-zero values to the
corresponding parameters in the input cards.

\section{ Usage of routines } Figure 2 shows an example of usage
of the CATCH routines.  The subroutine XPREP asks for a filename
 with a problem description, and performs some preliminary
procedures.  The subroutine XCATCH takes the entry values of $x$,
$\theta_x$, $y$, $\theta_y$ and $E$ and returns the exit values.
The entry/exit values of X, Y, XP, YP and ENERGY are double
precision .  The units in CALL...(..) are radians and metres; the
energy is returned modified according to the E loss in crystal;
hitting/missing the crystal is checked, the flag for that is
ISTATU (returned) :

\begin{itemize}
\item
 ISTATU = 0,    crystal missed,
\item
 ISTATU = 1, or --1, crystal hit,
\item
 ISTATU $<$ 0,  there was nuclear interaction.
\end{itemize}
In the case of interaction one may call the necessary routines to
produce the secondary particles and to get their exit parameters
modified for multiple scattering.  The XPOST can be called to save
the histograms, to fit the data, etc.  Figure 3 shows an example
of the input file.  The input data is listed below with comments
when necessary.
\begin{itemize}
\item
title
\item
KGEOM,   geometry type: either 110 (equidistant planes)
or 111 (planes like those in the Diamond(111) lattice)
\item
 KTRACE, trace level:
\begin{itemize}
\item
if $\le$ 0, no output during tracking
\item
if $=$ 1, some output at every step of tracking
\item
if $\ge$ 2, all current parameters are printed at every step \end{itemize}
\item
KRADV, switch on/off $R = R(z)$ option;
with KRADV = 1 one must supply FUNCTION CURVAT,
evaluating the curvature $1/R$ for every point;
with KRADV = 0 the curvature is 1/RADIUS.
\item
 KRADVY, switch on/off face twist;
with KRADVY=1 one must supply functions TCFROY and TCENDY,
evaluating the orientation of the planes for any point at
the crystal faces; with KRADVY = 0 that orientation is constant.
\item
 I1, I2, initial numbers for pseudo-random generator RAN(I1,I2)
 \item
 DZ, step size, microns. Try different DZ to be sure the result
 is independent of the step size.
A step of the order of 1 $\mu$m
 is suggested for 450 GeV/$c$ protons in Si(110).
The reasonable value of DZ should be
scaled like $\sqrt{E}$ (like the oscillation period in channelling)
\item
ENERGY, GeV
\item
PMASS, mass of the beam particle, (GeV)
\item
RADIUS, in cm, for the bent part of the crystal. When KRADV = 1
is set, the RADIUS is used to define the histogram windows only.
\item
DLINA, length of the bent part of the crystal  (cm).
The whole length of the crystal is STRAIT + DLINA. When KRADV = 0 is
set, the forthcoming part of the length STRAIT is assumed to be straight.
The following part of the th DLINA is assumed to be bent with R = RADIUS.
\item
STRAIT, length of the forthcoming straight part, (cm)
\item
 HDL, interplanar half-spacing, for (110) or (111)L, in \AA
 \item
HDS, interplanar half-spacing for (111)S, in \AA.
Must be 0 for KGEOM=110
\item
 U, thermal vibration amplitude of the lattice atoms,(\AA)
 \item
 DENS, crystal density, (gram/cm$^3$)
\item
 RDLI,  radia ion length, (cm)
\item
 ADLI, absorption length, (cm)
\item
 ZN,      charge of the nucleus
\item
 AN,      atomic weight
\item
 X0XTL,  the crystal entry-face left-edge $x$-coordinate, (cm)
 \item
  X1XTL,  the crystal entry-face right-edge $x$-coordinate, (cm)
   \item
  Y0XTL,  the crystal entry-face left-edge $y$-coordinate, (cm)
  \item
  Y1XTL,  the crystal entry-face right-edge $y$-coordinate, (cm)
   \item
THXTAL, crystal angle $\theta_x$, ($\mu$rad)
\item
THYTAL, crystal angle $\theta_y$, ($\mu$rad)
\item
SKINTH, amorphous skin thickness, (cm)\\
Some amorphous skin with uniform thickness SKINTH
 can be  superimposed over the  surface.
\item
ROUGH, (cm)
the roughness (= non-flatness) amplitude of every surface.
It is described by some periodical function, therefore you
should supply in the input both its amplitude ROUGH and period ROUGPD.
It is assumed that material under
the rough surface is crystalline.
\item
ROUGPD, surface roughness period, (cm)
\item
SKIMOZ, ($\mu$rad)
 There is the same skin, SKINTH + ROUGH,  at the entry face.
 If you wish, you can also set some angular spread SKIMOZ
 of the crystal lattice in the entry-face skin (in this skin only!).
The SKIMOZ means the sigma of Gaussian distribution.
\item
DISCUT, miscut angle, ($\mu$rad)
Note that its sign is essential;
when your impact is near the left edge of the crystal,
the negative sign (DISCUT$ <$0) is preferable.
\end{itemize}
The position of all surfaces in space, as a function of
$\theta_x$ and $\theta_y$ misalignment of the crystal
(THXTAL and THYTAL in the input cards), as a
function of bending with variable
radius, as well as a function of the surface  'rough structure',
is computed at every step.

All variables read from the input cards are placed into COMMON
blocks, and hence you can vary them from the main program during
the execution (if you wish to scan or to optimize).  Every
particle (both channelled and non-channelled) is tracked in the
crystal (X, Y are changed) and can leak out through any surface.
Near the rough surface it is even possible to leak out and be
caught again many times, i.e. the particle traverses sequentially
 the crystal bumps and vacuum (or amorphous skin) between. Hence,
 for particles touching  any surface one has both Coulomb multiple
and coherent collisions.

\section{ Acknowledgements } The  hospitality of the
CERN SL Division, in particular of W.~Scandale, and discussions
with users of CATCH at Protvino, Pisa, and CERN have fruitfully
contributed to this work.

\pagebreak

\pagebreak

\begin{figure} {
\small
\tt

\begin{verbatim}
PROGRAM XMAIN
C
C------------------------------------------------------------
C
An EXAMPLE of the
C  Monte Carlo simulation
C  of the planar channelling in bent crystals
C------------------------------------------------------------
DOUBLE PRECISION X,Y,XP,YP,ENERGY
C
C ...
   CALL XPREP            ! preliminary procedures
C ...
   NSTAT = 100
C               statistics cycle up to NSTAT events
C ...
   DO 40    IS = 1, NSTAT
C ...
   XP = GAUS (0.,20.E-6)
   X  = 2.E-3+1.E-6*RAN(I1,I2)
   YP = 0.0
   Y  = 0.0
   ENERGY  = 120.
C ...
  CALL XCATCH (X,XP,Y,YP,ISTATU,ENERGY)   ! crystal simulation
C ...
   if(ISTATU.LT.0) TYPE *,'interacted'
C ...
   TYPE *,IS,  XP,YP,  ISTATU
C ...
  40	CONTINUE              ! end of cycle
C ...
   CALL XPOST            ! some output
C ...
C =========================================================
   STOP
   END
\end{verbatim} }

\caption{An example of usage of the CATCH routines }
\end{figure}

\pagebreak
\begin{figure}{
\small
\tt
\begin{verbatim}
 LHC-EXTRACTION  simulation by CATCH $KEYS
  KGEOM  = 110, ! geometry type
  KTRACE = 0, ! trace level
  KRADV  = 0, ! switch on/off R = R(x,y,z) option
  KRADVY = 0, ! switch on/off face twist
 $END
 $OPTN
  I1 =  84837,  ! pseudo-random start
  I2 =  53463,  !  -//-
  DZ =  5., ! step size, microns
 $END
 $BEAM
  ENERGY = 7700., !  GeV
  PMASS  = .938,  ! mass of particle, GeV
 $END
 $CRYS
  RADIUS = 1.e4,  !  cm
  DLINA  = 7., ! length of bent part, cm
  STRAIT  = 0., ! length of forthcoming straight part, cm
  HDL =   .96, ! interplanar half-spacing, (110) or (111)L
  HDS =   .0, ! interplanar half-spacing, (111)S, Angst.
  U   =   .075,  ! thermal vibration amplitude, Angstroems
  DENS =   2.33,  ! density, gramm/cm3
  RDLI =   9.38,  ! radiation length, cm
  ADLI =   45., ! absorbtion length, cm
  ZN =   14., ! charge of nucleus
  AN =   28.09, ! atomic weight
 $END
 $GEOM
  X0XTL  = 0.2, ! min X of the Xtal-face (cm)
  X1XTL  = 0.5, ! max X of the Xtal-face (cm)
  Y0XTL  = -.15,! min Y of the Xtal-face (cm)
  Y1XTL  = .15, ! max Y of the Xtal-face (cm)
  THXTAL = .0, ! xtal angle wrt X (microrad)
  THYTAL = .0, ! xtal angle wrt Y (microrad)
  SKINTH = 0.E-5, ! amorphous skin thickness (cm)
  ROUGH  = 0.E-4, ! surface roughness (cm)
  ROUGPD = .01, ! surface roughness period (cm)
 SKIMOZ  = 10., ! skin mozaicity (microrad)
 DISCUT  = 0.,  ! miscut angle (microrad)
 $END
\end{verbatim}   }
\caption{An example of the input file }
\end{figure}
\end{document}